\documentclass[pra,twocolumn,showpacs,floats,eqsecnum,amsmath,amssymb]{revtex4}

\def\al{&\!\!\!\!}

\begin{document}

\title { Harmonic Oscillators and Elementary Particles}
\author {Y. Sobouti}
\affiliation{Institute for Advanced Studies in Basic Sciences (IASBS)\\
 No. 444, Prof. Sobouti Blvd., Zanjan 45137-66731, Iran \\
     email: sobouti@iasbs.ac.ir}

\begin{abstract}

Two dynamical systems with same  symmetry should have features in common,  and as far as their shared  symmetry is concerned, one may represent the other. 
The three light quark constituents of the hadrons,  a) have an   approximate flavor SU($3_f$) symmetry, b) have an  exact color-SU($3_c$) symmetry, and c) as spin $ \frac{1}{2} $  particles, have a  Lorentz SO(3,1) symmetry. So does  a 3D harmonic oscillator.   a) Its Hamiltonian has the  SU(3) symmetry, breakable if  the 3 fundamental modes of oscillation are not identical.  b) The 3 directions of oscillation have the permutation symmetry. This  enables one to create three copies of unbreakable SU(3) symmetry for each mode of the oscillation, and to mimic the color  of the elementary particles. And c) The Lagrangian of the 3D oscillator has the SO(3,1) symmetry.  This can be employed to accommodate the spin of the particles.   
  In this paper we  draw up a one-to-one correspondence between the eigen modes of the Poisson bracket operator of the  3D oscillator and  the flavor multiplets of the particles, and between the permuted modes of the oscillator and the color and anticolor multiplets of the particles. Gluons are represented by the  generators of the color SU($3_c$) symmetry of the oscillator.  
  
  Harmonic oscillators are common place objects and, wherever encountered, are analytically solvable. Elementary particles, on the other hand,  are abstract entities far from one's reach. Understanding of one may help a better appreciation of the other.

\end{abstract}


\maketitle
\section{Introduction}\label{introduction}

Jordan, 1935, is the initiator of the map from matrices to quantum harmonic oscillators to expedite computation with Lie algebra representations \cite{jordan}. Schwinger, 1952, evidently unaware of Jordan's work, represents the SU(2) algebra of the  angular momentum by two uncoupled quantum oscillators  \cite{schwinger}. Since then an extensive literature is created on the subject. The technique often bears the name of `Jordan-Schwinger map'.  In the majority of the existing literature,  the oscillator is a quantum one.   In their stellar system studies, however,  Sobouti et al \cite{sobouti}  and  \cite{dehghani}  associate the symmetries of their system of interest  with those of the  classical oscillators. They use  Poisson brackets  instead of the quantum commutation brackets, and work with complex functions in the  phase space of the oscillator.  Man'ko et al  do the same, and give a realization of the Lie product in terms of Poisson brackets \cite{manko}. 

In this paper we follow the classic oscillator approach  and explore the two-way association of SU(\textit{n}) $\leftrightarrows$ \textit{n}D oscillators. The case \textit{n}= 2, of course, gives the  the oscillator representation of the angular momentum, albeit in a different space and different notation than those of Schwinger.  The cases \textit{n}= 3, 4, ..., should be of relevance to particle physics. The flavor and color triplets of the three light quarks, (\textit{u, d, s}), and their higher multiplets have the SU(3) symmetry. They can be given a 3D oscillator representation. By inviting in the heavier quarks there might  also be room for higher SU(\textit{n}) and higher \textit{n}D  oscillators.

In his seminal paper Schwinger writes:
\textit{`... harmonic oscillator ...  provides a powerful method for constructing and developing the properties of angular momentum eigen vectors.  ... many known theorems  are derived in this way, and some new results obtained.'}
 Schwinger can only be right in saying so, for wherever encountered, harmonic oscillators are exactly  and easily solvable. It is in this spirit that we hope a harmonic representation of elementary particles might offer a simpler  and easier  understanding of at least the rudiments of the particle physics, if not lead to a different  insight. Classical harmonic oscillators are common place objects and can be set up on table tops. Elementary particles, on the other hand, are highly abstract notions and far from one's intuition.

\section{$n$D oscillators and their symmetries}
Let $(q_i, p_i; i=1, 2, ..., n)$ be the canonically conjugate pairs of coordinates and momenta of an $n$ dimensional ($n$D) harmonic oscillator, or equivalently of $n$ uncoupled oscillators. The Hamiltonian and  the Lagrangian are
$$H= \frac{1}{2}(p^2+ q^2), ~~~~~L = \frac{1}{2}(p^2- q^2), $$
 where  $p^2 =   p_ip_i, ... $. 
The time evolution of an attribute of the oscillator,  a function $f(p_i(t), q_i(t), t);~~i=1,...,n$ say, on the phase space trajectory of the oscillator is governed by Liouville's equation, 
\begin{equation}
i\frac{\partial f}{\partial t} = - i [f,H]_{poisson} = - i \left( p_i \frac{\partial}{\partial q_i} - q_i \frac{\partial}{\partial p_i} \right)f =: {\cal L} f \label{liouville}.
\end{equation}
The last equality is the definition of the the Poisson bracket operator, $\cal L $.  Hereafter, it will be referred to as  \textit{Liouville's operator}.  The reason for multiplication by $i$ is to render $\cal L$ hermitian and talk of its eigen solutions.  It should be noted that $\cal L$ is the sum of $n$ linear first order differential operators  ${\cal L} = {\cal L}_1 +...+ {\cal L}_n$. They are independent. Each  ${\cal L}_i$ depends only on the canonical coordinate-momentum pair, $(p_i, q_i)$, with no interaction with the other pairs.

\subsection{The Symmetries of ${\cal L}$, $H$, and $L$}\label{symliouville}
The most general infinitesimal coordinate transformation that leaves $\cal L$ invariant is the following,
\begin{eqnarray}
q'_i ~\al=\al ~q_i + \epsilon \left(   a_{ij} q_j + b_{ij} p_j\right),  \nonumber\\
p'_i ~\al=\al ~p_i + \epsilon \left( - b_{ij} q_j + a_{ij} p_j\right), ~\epsilon ~\text{infinitesimal},\label{inftesimal}
\end{eqnarray}
The transformation is linear, and  $a=[a_{ij}]$ and  $b=[b_{ij}]$ are two $n\times n$ matrices. The proof for $n=3$ is given in   \cite{sobouti} and   \cite{dehghani}. Its generalization  to higher dimensions is a matter of letting the subscripts $i$ and $j$ in Eqs. (\ref{liouville}) and (\ref{inftesimal}) span the range 1 to $n$.  There are $2n^2$ ways to choose  the $a$- and $b$- matrices, showing that the symmetry group of $\cal L$ and thereof that of Eq.(\ref{liouville}) is GL(\textit{n,c}), the group of general $n\times n$ complex matrices.

At this stage let  us introduce  $\cal H$ as the function space  of all complex valued and square integrable functions, $f(p_i, q_i)$, in which the inner product is defined as
\begin{eqnarray}
(g,h)&=&\int g^*f\exp(-2E)d^np d^nq < \infty, ~~f, g\in {\cal H},\nonumber
\end{eqnarray}
where $E=\frac{1}{2}(p^2 + q^2)$ is the energy scalar of the Hamiltonian operator of the oscillator.
Associated with the transformation of Eq. (\ref{inftesimal}), are the following generators on the function space $\cal H$,
 \begin{eqnarray}
& \chi & = a_{jk} \left(  p_j \frac{\partial}{\partial p_k}  + q_j \frac{\partial}{\partial q_k}\right)
           - i b_{jk} \left(  p_j \frac{\partial}{\partial q_k} - q_j \frac{\partial}{\partial p_k} \right) , \nonumber\\       \label{generators}
 \end{eqnarray}
 (insertion of $-i$ in front of $b_{jk}$ is for later convenience). 
Again there are $2n^2$ generators.  All $\chi$'s  commute with $\cal L$ but not necessarily among themselves. 

Two notable subgroups of GL(\textit{n,c}) are generated by
\begin{itemize}
\item[1)]   $a_{ij}$  antisymmetric,  $b_{ij}$ antisymmetric,
\item[2)]  $a_{ij}$  antisymmetric,  $b_{ij}$ symmetric,
\end{itemize}
Case 1 is the symmetry group of the Lagrangian. It is  of Lorentz type, and in the 3D case reduces to SO(3,1),  the symmetry group of Minkowsky's spacetime and of Dirac's equation for spin $\frac{1}{2}$ particles. We will come back to it briefly in the conclusion of this paper. 
Case 2, antisymmetric $a_{ij}$  and symmetric  $b_{ij}$, is the symmetry of the Hamiltonian,  the SU($n$) group. 
Before proceeding further, however, let us give the proof of  the last two  statements.     Under the infinitesimal transformation of Eq. (\ref{inftesimal}) one has
\begin{eqnarray}
\delta L &=& \epsilon [a_{ij}(p_i p_j - q_i q_j) - b_{ij}(p_i q_j + q_i p_j)], \nonumber\\
\delta H &=& \epsilon [a_{ij}(p_i p_j + q_i q_j) - b_{ij}(p_i q_j - q_i p_j)]. \nonumber
\end{eqnarray} 
There follows
\begin{eqnarray}
\delta L &=& 0 \textrm{    if    } a_{ij}= - a_{ji},~\textrm{   and   }~ b_{ij}= - b_{ji}, \nonumber\\
\delta H &=& 0 \textrm{    if    } a_{ij}= - a_{ji},~\textrm{   and   }~ b_{ij}=b_{ji}. QED\nonumber
\end{eqnarray}
Comming back, an SU($n$) is spanned by $n^2-1$ linearly independent basis matrices. A convenient and commonly used basis for SU(\textit{n}) is the generalized Gell-Mann's       $\lambda$ matrices. See e.g. \cite{greiner} for their construction and see Table \ref{gellmann matrices} for a refresher. The generalized $\lambda$ matrices consist of $\frac{1}{2}n(n-1)$ antisymmetric and imaginary matrices plus $\frac{1}{2}n(n+1)-1$ symmetric, real, and traceless ones. Their commutation brackets are:
\begin{equation}
\left[\frac{\lambda_a}{2}, \frac{\lambda_b}{2}\right] = i f_{abc}  \frac{\lambda_c}{2}, ~~a,b,c = 1, 2, ..., n^2 -1,\label{lambdaalgebra}
\end{equation}
where $ f_{abc}$ are the structure constants. All  are real and completely antisymmetric in $a, b, c.$
 To extract the corresponding $\chi$ generators from those of Eq. (\ref{generators}), we choose
\begin{eqnarray}
 a &=& \frac{1}{2} \lambda_{antisym.} = \frac{1}{2}(\lambda^2,   \lambda^5, \lambda^7, ...), \nonumber \\
 b &=& \frac{1}{2} \lambda_{sym} = \frac{1}{2}(\lambda^1, \lambda^3,  \lambda^4, \lambda^6, \lambda^8, ...).\nonumber
 \end{eqnarray}
 This choice produces a set of  $(n^2-1)$ linear differential operators, that are the oscillator representations of the  SU($n$) algebra in $\cal H$.
The first  8 of them are as follows:
 \begin{eqnarray}
 \chi_1 &=& -\frac{i}{2}\left\{\left( p_1\frac{\partial}{\partial q_2} - q_1\frac{\partial}{\partial p_2}\right)
                         + \left(p_2\frac{\partial}{\partial q_1} - q_2\frac{\partial}{\partial p_1}\right) \right\},\nonumber\\
  \chi_2 &=& - \frac{i}{2}\left\{\left(p_1\frac{\partial}{\partial p_2}+q_1\frac{\partial}{\partial q_2}\right)
                                       -\left(  p_2\frac{\partial}{\partial p_1}   + q_2\frac{\partial}{\partial q_1}\right) \right\},\nonumber\\
 \chi_3  &=& - \frac{i}{2}\left\{\left( p_1\frac{\partial}{\partial q_1} - q_1\frac{\partial}{\partial p_1}\right)
                         -  \left( p_2\frac{\partial}{\partial q_2} - q_2\frac{\partial}{\partial p_2}\right) \right\}\nonumber\\
             &=&\frac{1}{2} ({\cal L}_1 - {\cal L}_2).     \nonumber\\
 \chi_4  &=& -\frac{i}{2}\left\{\left( p_1\frac{\partial}{\partial q_3} - q_1\frac{\partial}{\partial p_3}\right)
                         + \left(p_3\frac{\partial}{\partial q_1} - q_3\frac{\partial}{\partial p_1}\right) \right\},\nonumber\\
  \chi_5 &=& -\frac{i}{2}\left\{\left( p_1\frac{\partial}{\partial p_3} + q_1\frac{\partial}{\partial q_3} \right)
                      - \left( p_3\frac{\partial}{\partial p_1} + q_3\frac{\partial}{\partial q_1}\right) \right\},\nonumber\\
   \chi_6 &=& -\frac{i}{2}\left\{\left( p_2\frac{\partial}{\partial q_3} - q_2\frac{\partial}{\partial p_3}\right)
                         + \left(p_3\frac{\partial}{\partial q_2} - q_3\frac{\partial}{\partial p_2}\right) \right\},\nonumber\\
    \chi_7 &=& -\frac{i}{2} \left\{\left(p_2\frac{\partial}{\partial p_3} +q_2\frac{\partial}{\partial q_3}\right)
                                - \left(  p_3\frac{\partial}{\partial p_2} +q_3\frac{\partial}{\partial q_2}\right) \right\},\nonumber\\
      &~&\nonumber\\
  \chi_8 &=& -\frac{i}{2\sqrt{3}}\left\{\left(p_1\frac{\partial}{\partial q_1} - q_1\frac{\partial}{\partial p_1}\right)
                                 +      \left(p_2\frac{\partial}{\partial q_2} - q_2\frac{\partial}{\partial p_2}\right)\right\}
                                  \nonumber\\
            &+&  \frac{i}{\sqrt{3}}\left\{\left(p_3\frac{\partial}{\partial q_3} - q_3\frac{\partial}{\partial p_3} \right) \right\}  \nonumber\\
          &=& \frac{1}{2\sqrt{3}}({\cal L}_1 + {\cal L}_2 - 2{\cal L}_3).   \label{fgenerators}
 \end{eqnarray}
All $\chi$'s are hermitian.
Their commutation brackets  are the same as those of $\lambda$'s,
$$ [\chi_a,\chi_b] =i f_{abc} \chi_c .$$

\subsection{Solutions of Equation (\ref{liouville})}

The followings can be easily verified
\begin{eqnarray}
&~&{\cal L} (p_i \pm i q_i)^n~=~ \pm n(p_i\pm iq_i)^n, \label{rudiment}\\
&~&{\cal L} E = 0, ~~ {\cal L}\exp(-E) = 0, ~~E = \frac{1}{2}(p^2 + q^2).\nonumber 
\end{eqnarray}
Let us use the shorthand notation  $ z_i = p_i + i q_i$ and $E=z_i z^*_i/2$. Considering the fact that Eq. (\ref{liouville}) is a linear differential equation,  and
${\cal L}={\cal L}_1+{\cal L}_2+...$ is the sum of $n$ independent operators, one immediately writes down the (unnormalized) eigen states  of Eq. (\ref{liouville}) as:
\begin{eqnarray}
  f^{m_1 ...  m_n}_{n_1 ... n_n}
 \al=\al z_1^{n_1} ... z_n^{n_n} z_1^{*m_1} ...  z_n^{*m_n}
\exp[-i t (n-m)], \nonumber\\
\al~\al \nonumber\\
{\cal L} f^{m_1 ... m_n}_{n_1 ... n_n} \al=\al (n-m) f^{m_1 ... m_n}_{n_1 ... n_n}, \label{eigenval}
\end{eqnarray}

\noindent
where
$ n=\sum_i n_i, ~ m=\sum_i m_i$.
The modes reported in Eq.  (\ref{eigenval}) are members of the function space $\cal H$. As defined in  Eq. (\ref{inproduct}), the explicit form of their inner product is
\begin{eqnarray}
&&( f^{m'_1 ...  m'_n}_{n'_1 ... n'_n},  f^{m_1 ...  m_n}_{n_1 ... n_n}) \nonumber\\
&&~~~~~= \int z_1^{n_1+m'_1}...  z_n^{n_n+m'_n} z_1^{*n'_1+m_1}...  z_n^{*n'_n+m_n} ]\nonumber\\
&&\hspace{3.5cm}\times \exp(-2E) d^n p d^n q \nonumber\\
 &&~~~~~\propto  \delta_{(n_1+m'_1), (n'_1+ m_1)}...\delta_{(n_1+m'_1), (n'_n+ m_n)}. \label{inproduct}
\end{eqnarray}

\noindent
The eigenvalue $(n-m)$ of Eq. (\ref{eigenval})is degenerate. Different combinations of $n_i$ and $m_i$ can give the same $n-m$.
Degenerate sets of modes are not in general  orthogonal ones.
However, for each multiplet of given $n$ and $m$  it is always possible to construct   an orthonormal set through  suitable  linear combinations of  the multiplet members.

So much for generalities. In section \ref{angmom} we examine the 2D oscillator representation of SU(2) and reproduce Schwinger's model, albeit in a different notation. Next, we treat the 3D case and suggest a scheme, that we think, it represents the  flavor and color symmetries of the light quarks and higher particle multiplets.

\section{2D Oscillator and  SU(2) - Angular Momentum}\label{angmom}

There are two complex planes to deal with, $z_1=p_1+iq_1$ and $z_2=p_2+iq_2$.  
The first three operators of Eqs. (\ref{fgenerators}) are the  ones to work with,  and have  the SU(2) algebra,
\begin{equation}
[\chi_i,\chi_j]= i \epsilon_{ijk}\chi_k;  \hspace{0.2cm} i,j,k = 1,2,3. \label{su2gen}
\end{equation}
Therefore, one may construct the following familiar angular momentum operators and commutators 
\begin{eqnarray}
&&J_i = \chi_i,~~~~J_\pm = J_1 \pm i J_2, \label{jpm}\\
&&J^2 = J_1^2 + J_2^2 + J_3^2 = J_+ J_- + J_3^2 - J_3.   \label{jsquared}\\
&&[J^2, J_i] = 0,    ~~~~ [J_z, J_\pm] = \pm J_\pm. \label{jcommutators}
\end{eqnarray}
The common eigen-states of $(J^2, J_3)$ (in ket notation) should be of the form
\begin{eqnarray}
&& J^2 |j, m> = j(j+1) |j, m>, ~~~\nonumber\\
&&J_3 |j, m> = m |j, m>, ~~~~-j \leq m \leq j.\nonumber
\end{eqnarray}
In Table \ref{chi-on-z} we have collected the outcome of the operation of $\chi_i$ on $z_j= p_j + iq_j$. One may readily verify that
\begin{eqnarray}
&& |\frac{1}{2}, \frac{1}{2}> = z_1,~~~ |\frac{1}{2}, -\frac{1}{2}> = z_2,\nonumber\\
&& |1,1> = z_1^2,  ~~~ |1, 0> = z_1z_2, ~~~ |1,-1> = z_2^2.\nonumber
\end{eqnarray}
The general rule is
$$ |j, m> = z_1^{j+m} z_2^{j-m}.$$
In particular,
$$ |j, j> = z_1^{2j}, ~~~|j,0> = z_1^j z_2^j, ~~~ |j, -j> = z_2^{2j}. $$
One may, of course, begin with any $|j,m>$ and reach the other $(2j+1)$ members of the $j$-multiplet by operating on   $|j,m>$ with the raising and lowering ladders  $ J_\pm$.

\section{3D oscillator and  Quark Flavor}\label{flavor}

There are three complex planes to deal with, 
$$z_1=p_1+iq_1,~  z_2=p_2+iq_2, \textrm{~ and~ } z_3=p_2+iq_3.$$
 The eight generators of Eq. (\ref{fgenerators}) are the relevant ones. Two  of them,
$\chi_3 $ and $\chi_8,$ commute together and commute with ${\cal L} = {\cal L}_1+ {\cal L}_2 + {\cal L}_3.$ From them (using the particle physics nomenclature) we  compose the following linear combinations,
 \begin{eqnarray}
 I_3 &=& \chi_3 = \frac{1}{2}({\cal L}_1 - {\cal L}_2),~~~~~~~~~~~~~~~~ \textrm{ isospin,}\nonumber\\
 Y &=&   \frac{2}{\sqrt{3}} \chi_8 = \frac{1}{3}({\cal L}_1 + {\cal L}_2 - 2 {\cal L}_3 ), \textrm{   hypercharge,} \nonumber\\
 B &=&   \frac{1}{3}( {\cal L}_1 + {\cal L}_2 +{\cal L}_3) , ~~~~~ \textrm{ baryon number,} \nonumber \\
 Q &=& I_3+\frac{1}{2} Y = \frac{1}{3}(2{\cal L}_1 - {\cal L}_2 - {\cal L}_3) , \textrm{ charge,}\nonumber\\
  S &=& Y-B = -{\cal L}_3,   ~~~~~~~~~~~~~~~ \textrm{strangeness,}\label{i3yb} 
 \end{eqnarray}
Only three of these five operators are linearly independent. Their eigenvalues and eigen-states can be read from Eqs. (\ref{eigenval}),
\begin{eqnarray}
I_3 f_{n_1n_2n_3}^{m_1m_2m_3} &=& \frac{1}{2} [(n_1-m_1) - (n_2-m_2)]  f_{n_1n_2n_3}^{m_1m_2m_3}\nonumber\\
Y f_{n_1n_2n_3}^{m_1m_2m_3} &=& \frac{1}{3} [(n_1-m_1)
 + (n_2-m_2)\nonumber\\
 &~&\hspace{1.8 cm}  - 2(n_3-m_3]  f_{n_1n_2n_3}^{m_1m_2m_3} \nonumber\\
B f_{n_1n_2n_3}^{m_1m_2m_3} &=& \frac{1}{3} [(n_1-m_1) + (n_2-m_2) \nonumber\\
&~&\hspace{1.8 cm} + (n_3-m_3]  f_{n_1n_2n_3}^{m_1m_2m_3} \nonumber\\
Q f_{n_1n_2n_3}^{m_1m_2m_3} &=& \frac{1}{3} [2(n_1-m_1)  - (n_2-m_2) \nonumber\\
&~&\hspace{1.8 cm} - (n_3-m_3]  f_{n_1n_2n_3}^{m_1m_2m_3}  \nonumber\\
\nonumber\\
S f_{n_1n_2n_3}^{m_1m_2m_3} &=& -(n_3-m_3)  f_{n_1n_2n_3}^{m_1m_2m_3}\nonumber\\ \label{i3ybqs}
\end{eqnarray}

The collection of all eigen-states belonging to a given $n= n_1+n_2+n_3$ and $m= m_1+m_2+m_3$ will be called a multiplet and will be denoted by D($n,m$). For instance, D(1.0) will be the  triplet of $ f_{100}^{000}=z_1,  f_{010}^{000}=z_2,   f_{001}^{000}=z_3$. Below we examine some of these multiplets, compare their eigen characteristics with those of the known particle multiplets, and point out  the one-to-one correspondence between the two.   \\

Table \ref{Table:Quark and Antiquark}   displays the two triplets D(1, 0) and D(0, 1). The first of which consists of $ (z_1, z_2, z_3) $ and the second of  $(z^*_1, z^*_2, z^*_3).$ Their baryon, isospin, hypercharge, charge, and strangeness numbers, are read from  Eqs. (\ref{i3ybqs}).  They are the same as those of the quark and antiquark triplets.  The members of the multiplet are orthogonal to one another in the sense of Eq. (\ref{inproduct}). The last column in  this and the other tables below displays  the Casimir numbers, $4I^2+3Y^2$, an index to identify the submultiplets within a multiplet.\\

Table \ref{Table: Pseudoscalar Meson Nonet} is D(1, 1). It has  nine members, ($z_i z^*_j;~ i,j = 1,2,3$), or their linear combinations.
D(1,1)  is identified with the pseudo-scalar meson nonet. There are two submultiplets to it, characterized by the two Casimir numbers 4 and 0. Again the 9 members of the multiplet constitute  a complete orthogonal set in the subspace of the  pseudo-scalar mesons.\\

Table \ref{Table: Baryon Decuplet} is D(3, 0) and is identified with the Baryon Decuplet.  It has ten members,
$z_1^i z_2^j z_3^k;~ i, j, k = 0,1,2,3$, constrained to $i+j+k=3.$ There are two submultiplet in the baryon decuplet, characterized by the two Casimir numbers 4 and 12. The ten members constitute a complete orthogonal set. 
The Antibaryon Decuplet is D(0, 3). It can be read from Table III by  simply interchanging the subscripts and superscripts in the first column of the Table and changing the signs of the eigenvalues accordingly. This also means interchanging $z_j\rightleftarrows z^*_j$ in the Table.\\

\section{Color and color multiplets}\label{color}

By mid 1960 particle physicists had felt the  need for an extra quantum number for quarks in order to comply with Pauli's exclusion principle and to justify coexistence of the like spin-$\frac{1}{2}$ quark flavors in baryons. The notion of  color and color charge was introduced; see e.g. \cite{greenberg}, \cite{han}, \cite{nambu}, and \cite{matveev}. The  consensus of opinion nowadays is that each quark flavor comes in three colors, \textit{red}, \textit{green} and \textit{blue};  and atinquarks in three anticolors, \textit{antired}, \textit{antigreen} and \textit{antiblue}. Strong interactions are mediated by 8 bicolored gluons, each carrying one color charge  and one anticolor charge. Color is believed to be conserved in the course of strong interactions. 
 
Does  the 3D harmonic oscillator has an attribute analogous to the color of the quarks, is that attribute conserved, and if so, what is the symmetry responsible for its conservation?  What are the counterparts of gluons in the 3D oscillator? 
In section \ref{symliouville} we talked about the continuous symmetries of the Hamiltonian and of the relevant Liouville equation  and  came up with a  one-to-one correspondence between the fundamental eigen modes of the oscillator and the quantum numbers of the quark flavors. There are discrete symmetries  to consider:
\begin{itemize}
 \item[A.]
  The Hamiltonian of a 3D oscillator  is symmetric and thereof  its  Liouville operator is antisymmetric under the discrete transformation
\begin{equation}
q_i \rightarrow p_i,~~~p_i \rightarrow q_i.\label{A}
\end{equation}
This leads to
 $${({{\cal L}z_i}= z_i)}^*\rightarrow{\cal L}z^*_i = - z^*_i, $$
 that is, the fact that $z_i$ and $z^*_i$ are the eigen states of $\cal L$ with eigenvalues $\pm 1$ is due to the symmetry of the Hamiltonian and the antisymmetry of $\cal L$ under the transformation of Eq. (\ref{A}). In particle physics language, this  is akin to the statement that if a particle is a reality, so is its antiparticle.
\item[B.]
The total Hamiltonian and the total $ {\cal L} = {\cal L}_1 +  {\cal L}_2 + {\cal L}_3 $ are symmetric under the permutation of the three dimension subscripts (1, 2, 3).  
 To clarify the point let us, for the moment, instead of talking of 3D oscillators, talk of  three uncoupled oscillators; and  distinguish between the 3 coordinate directions in the ($q,  p$) spaces and the 3 oscillators  $(1, 2, 3)$ we choose to assign to those directions.  Let us rename the three directions in the ($q, p$) spaces  as three boxes colored red(\textit{r}), green(\textit{g}), and blue(\textit{b})
 (the language we adopt is in the anticipation of finding a correspondence between the permutation symmetry of the 3D oscillator and the color symmetry of the quark triplets).
   One has the option to place any of the oscillators in any of the colored boxes, \textit{r}, \textit{g}, or  \textit{b}. The choices are:
   \begin{center}
$ r(1),  g(2), b(3)$\\
$ r(3),  g(1), b(2)$\\
$ r(2),  g(3), b(1)$
\end{center}
There are three copies of the oscillator 1: \textit{r}(1), \textit{g}(1), and  \textit{b}(1),  and similarly for the oscillators 2 and 3. Each oscillator can be in a triplet color state. Defined as such, the  color symmetry is exact and is not \textit{breakable}, in contrast to the flavor symmetry in which one assumes the 3 oscillators are identical, while it may be broken by allowing the masses and spring constants of the oscillators to change. 

As noted earlier the transformation of Eq. (\ref{A}),  $q_i \rightarrow p_i,~~~p_i \rightarrow q_i$,  amounts to going from the complex $(p,q)$ plane to its complex conjugate, $(p,q)^*$, plane.  One may now commute the coordinates in this complex conjugated space and design  an anticolor scheme,
  antired$(\overline{r})$, antigreen$(\overline{g})$, and antiblue$(\overline{b})$, say. Thus,
  \begin{center}
$ \overline{r}(1),  \overline{g}(2), \overline{b}(3)$\\
$ \overline{r}(3),  \overline{g}(1), \overline{b}(2)$\\
$ \overline{r}(2),  \overline{g}(3), \overline{b}(1)$\\
\end{center}
Again there are 3 copies of each oscillator in the  $(q,p)^*$ space, our analogue of the antiparticle domain. \end{itemize}

\section{Gluons}\label{gluon}

With the definition of the preceding section the \textit{color} is now a direction in the $(p, q)$ phase space. Each oscillator, while sharing the approximate flavor SU(3)$_{f}$ symmetry with the others, has its own exact fundamental triplet color SU(3)$_{c}$ symmetry. The \textit{adjoint  color} representation of SU(3)$_{c}$  is the same as the $\chi$ octet  of Eqs. (\ref{fgenerators}) in which the subscripts 1, 2, and 3 of $p$'s and $q$'s are now replaced by $r$, $g$, and $b$, respectively. Thus,   
\begin{eqnarray} 
\chi_1 &=& -\frac{i}{2}\left\{\left( p_r\frac{\partial}{\partial q_g} - q_r\frac{\partial}{\partial p_g}\right)
+ \left(p_g\frac{\partial}{\partial q_r} - q_g\frac{\partial}{\partial p_r}\right) \right\},\nonumber\\
\chi_2 &=& - \frac{i}{2}\left\{\left(p_r\frac{\partial}{\partial p_g}+q_r\frac{\partial}{\partial q_g}\right)
-\left( p_g\frac{\partial}{\partial p_r} + q_g\frac{\partial}{\partial q_r}\right) \right\},\nonumber\\
\chi_3 &=& - \frac{i}{2}\left\{\left( p_r\frac{\partial}{\partial q_r} - q_r\frac{\partial}{\partial p_r}\right)
- \left( p_g\frac{\partial}{\partial q_g} - q_g\frac{\partial}{\partial p_g}\right) \right\}\nonumber\\
&=&\frac{1}{2} ({\cal L}_r - {\cal L}_g). \nonumber\\
\chi_4 &=& -\frac{i}{2}\left\{\left( p_r\frac{\partial}{\partial q_b} - q_r\frac{\partial}{\partial p_b}\right)
+ \left(p_b\frac{\partial}{\partial q_r} - q_b\frac{\partial}{\partial p_r}\right) \right\},\nonumber\\
\chi_5 &=& -\frac{i}{2}\left\{\left( p_r\frac{\partial}{\partial p_b} + q_r\frac{\partial}{\partial q_b} \right)
- \left( p_b\frac{\partial}{\partial p_r} + q_b\frac{\partial}{\partial q_r}\right) \right\},\nonumber\\
\chi_6 &=& -\frac{i}{2}\left\{\left( p_g\frac{\partial}{\partial q_b} - q_g\frac{\partial}{\partial p_b}\right)
+ \left(p_b\frac{\partial}{\partial q_g} - q_b\frac{\partial}{\partial p_g}\right) \right\},\nonumber\\
\chi_7 &=& -\frac{i}{2} \left\{\left(p_g\frac{\partial}{\partial p_b} +q_g\frac{\partial}{\partial q_b}\right)
- \left( p_b\frac{\partial}{\partial p_g} +q_b\frac{\partial}{\partial q_g}\right) \right\},\nonumber\\
&~&\nonumber\\
\chi_8 &=& -\frac{i}{2\sqrt{3}}\left\{\left(p_r\frac{\partial}{\partial q_r} - q_r\frac{\partial}{\partial p_r}\right)
+ \left(p_g\frac{\partial}{\partial q_g} - q_g\frac{\partial}{\partial p_g}\right)\right\}
\nonumber\\
&+& \frac{i}{\sqrt{3}}\left\{\left(p_b\frac{\partial}{\partial q_b} - q_b\frac{\partial}{\partial p_b} \right) \right\} \nonumber\\
&=& \frac{1}{2\sqrt{3}}({\cal L}_r + {\cal L}_g - 2{\cal L}_b). \label{colorgenerators}
\end{eqnarray}
A typical differential operator, 
$$ \left(p_c\frac{\partial}{\partial q_{c'}} -q_ c\frac{\partial}{\partial p_{c'}}\right), ~ c, c' =r, g, b,$$
 in Eqs. (\ref{colorgenerators}),
upon operation on a typical color state $z_{c'}= p_{c'}+iq_{c'}$ annihilates the $c'$ color and create the $c$ color. 
Let us now use Table \ref{chi-on-z} and, as examples, see the action of $\chi_1$ and $\chi_2$ on the colored  oscillator states $z_r$ and $z_g$:
\begin{eqnarray}
\chi_1 z_r= \frac{1}{2} z_g, ~~~ &&\chi_1 z_g= \frac{1}{2} z_r,\nonumber\\
\chi_2 z_r= \frac{i}{2} z_g, ~~~ &&\chi_2 z_g=- \frac{i}{2} z_r,\nonumber
\end{eqnarray}
from which we obtain 
\begin{eqnarray}
(\chi_1 + i \chi_2) z_r = 0,   ~~~&& (\chi_1 + i \chi_2) z_g = z_r,\label{RGbar}\\
(\chi_1 - i \chi_2) z_r = z_g, ~~~&& (\chi_1 - i \chi_2) z_g = 0.\label{GRbar}
\end{eqnarray}
From Eq.(\ref{RGbar}) we learn that $ (\chi_1 + i\chi_2)$ annihilates a green state  and creates a red one. Shall we denote it by  $R\overline{G}$ and call it a gluon bicolored antigreen and red? 
Similarly, $(\chi_1 - i \chi_2)$ can be denoted by $G\overline{R}$ and called a green and antired gluon. 
Note that expressions such as $ R\overline{G}$ and $G\overline{R}$ are differential operators. Also note that terms such as $G\overline{R} z_g $ and $R\overline{G} z_r $ yield zero; for there is no red in green or green in red to be extracted and turned into another color state.

We are now in a position to draw up a gluon table in terms of $\chi$'s by generalizing the examples above.
\begin{eqnarray}
R\overline{G} &=& \chi_1 + i \chi_2, ~~~ \nonumber\\
G\overline{R} &=& \chi_1 - i \chi_2,\nonumber\\
R\overline{R} &-& G\overline{G} = 2\chi_3\nonumber\\
R\overline{B} &=& \chi_4 + i \chi_5, ~~~\nonumber\\
B\overline{R} &=& \chi_4 - i \chi_5, \nonumber \\
G\overline{B} &=&\chi_6 + i \chi_7, ~~~\nonumber\\
B\overline{G} &=& \chi_6 - i \chi_7, \nonumber\\
R\overline{R} &+& G\overline{G} -2 B\overline{B} = 2\sqrt{3} \chi_8. \label{gluon octet}
\end{eqnarray}
The inverse relations are
\begin{eqnarray}
\chi_1 &=& \frac{1}{2}(R\overline{G} + G\overline{R}), ~~~ \chi_2 = \frac{1}{2i}(R\overline{G} - G\overline{R}), \nonumber\\ 
\chi_3 &=& \frac{1}{2}(R\overline{R} - G\overline{G}), \nonumber\\
 \chi_4 &=& \frac{1}{2}(R\overline{B} + B\overline{R}), ~~~ \chi_5 = \frac{1}{2i}(R\overline{B} - B\overline{R}), \nonumber\\ 
 \chi_6 &=& \frac{1}{2}(G\overline{B} + B\overline{G}), ~~~ \chi_7 = \frac{1}{2i}(G\overline{B} - B\overline{G}), \nonumber\\ 
 \chi_8 &=& \frac{1}{2\sqrt{3}}(R\overline{R} + G\overline{G} - 2  B\overline{B}),  \label{gluon octet2}\\
 {\cal L} &=& {\cal L}_1 + {\cal L}_2 + {\cal L}_3 =  R\overline{R} + G\overline{G} +  B\overline{B}.\label{gluon singlet}
\end{eqnarray}
In  Eqs.(\ref{gluon octet}) and (\ref{gluon octet2}), $\chi_3$ and $\chi_8$ are colorless and members of a color octet. In Eqs. (\ref{gluon singlet}), ${\cal L}$ is colorless and a singlet.

 \subsection{Multiplication table of gluons}\label{gluon multiplication table}
 
 We begin with examples  
 \begin{eqnarray}
&1.& R\overline{G}(G\overline{B} z_b) =  R\overline{G} z_g = z_r. \textrm{  Thus,  } R\overline{G}G\overline{B} = R\overline{B}. 
\nonumber\\
&2.& R\overline{G}(B\overline{G} z_g) =  R\overline{G} z_r = 0. \textrm{  Thus,  } R\overline{G}B\overline{G} = 0.\nonumber
\end{eqnarray}
Evidently, in  multiplying several gluon operators, terms of the form $\overline{R}R,  \overline{G}G$ and    $\overline{B}B $ annihilate each other leaving a unit operator behind. Products of the form $\overline{R}G  \overline{G}B$, $\overline{G} B\overline{B}R $, and $\overline{B}R \overline{R}G$ yield zero. Guided by these examples we draw up the gluon multiplication Table \ref{glu multip table}.

\section{Concluding remarks}

Among the symmetries of an $n$D oscillator is the symmetry group  of its Hamiltonian,  SU($n$),  a subgroup of the symmetries of its Poisson bracket, GL($n, c$). This feature can be employed to represent any system with SU($n$) symmetry by an $n$D oscillator.  Schwinger's representation of the angular momentum by two uncoupled oscillators and our version in section \ref{su2gen}  are examples of such representations.  

In section \ref{flavor} we propose a 3D oscillator representation of the flavor of the elementary particles. We present the quark and antiquark triplets, the pseudoscalar meson nonet, and the baryon decuplet. These follow from the continuous symmetries of  Liouville's operator under simultaneous rotations of the   p- and q- axes, and rotations in  the (p,q)  planes;  see the roles of $a_{ij}$ and $b_{ij}$ in  Eqs. (\ref{inftesimal}).  

In section \ref{color} we deal with color symmetry. The Poisson bracket (the Liouville operator) of the  3D oscillator  is also symmetric under permutation of the 3 directions in the $(p, q)$ spaces; and antisymmetric under the exchange of $p\rightarrow q$ and $q\rightarrow p$.  These discrete symmetries combined with the continuous SU(3) symmetry of the permuted systems gives rise to a color-like SU(3) symmetry, the adjoint representation of which in the function space, $\cal H$, is a set of gluon-like operators, responsible for permuting the `color' of the oscillators (i.e. the color of the quarks).

It was mentioned  before, the symmetry group of the Lagrangian   of the 3D oscillator is SO(3,1).  The generators of this sub-algebra in oscillator representation can be found in \cite{sobouti} and \cite{dehghani}.   On the other hand, SO(3,1) is also the symmetry of  Minkowsky's spacetime and  of the  spin one-half Dirac particles. There is the possibility of employing this feature to give a 3D oscillators representation of the spin of the elementary particles. We have not, however, followed this line of thought here to its conclusion.

A pedagogical note; the $n$D oscillator introduced here is  a classical one; no quantum feature is attached to it. Yet this classical system is capable of representing some aspects of the elementary particles, highly abstract quantum mechanical and  field theoretical notions. Classical systems can be constructed on table tops.  One wonders whether it is possible to demonstrate, at least the rudiments,  of the elementary particles by some oscillator-based devices. For example to show that a 3D mechanical oscillator cannot escape to infinity. It is confined to the vicinity of the origin, and the  closer to the origin it stays  the freer it is;  a way to convey the quark  confinement and asymptotic freedom? Or whether the characteristics of the Lissajous type oscillation modes of the oscillator has any feature in common with particles, or whether the Lissajous mode can represent any of the particles?  

\textbf{Acknowledgement:} The author wishes to thank M B Jahani Poshteh for reading several pre- publication versions of this paper and for offering valuable suggestions.

\begin{table*}[ht]
\caption{Gell-Mann Matrices}
\vspace{0.3 cm}
\begin{tabular}{c  c  c}

\hline\\
$\lambda_1 = \left[ \begin{array}{c c c} 0&1&0\\ 1&0&0\\  0&0&0\\ \end{array} \right],$ &
$\lambda_2 = \left[ \begin{array}{c c c} 0&-i&0\\ i&0&0\\  0&0&0\\ \end{array} \right],$ &
$\lambda_3 = \left[ \begin{array}{c c c} 1&0&0\\ 0&-1&0\\  0&0&0\\ \end{array}  \right],$\\
\\
$\lambda_4 = \left[ \begin{array}{c c c} 0&0&1\\ 0&0&0\\  1&0&0\\ \end{array} \right],$ &
$\lambda_5 = \left[ \begin{array}{c c c} 0&0&-i\\ 0&o&0\\  i&0&0\\ \end{array} \right], $&
$\lambda_6 = \left[ \begin{array}{c c c} 0&0&0\\ 0&0&1\\  0&1&0\\ \end{array}\right], $\\
\\
$\lambda_7 = \left[ \begin{array}{c c c} 0&0&0\\ 0&0&-i\\  0&i&0\\ \end{array} \right],$ &
$~~~~~\lambda_8 = \frac{1}{\sqrt{3}}\left[ \begin{array}{c c c} 1&0&0\\ 0&1&0\\  0&0&-2\\ \end{array} \right].$ &

\label{gellmann matrices}
\end{tabular}

\end{table*}

\begin{table*}[ht]
\caption{$\chi_i z_j$ and $\chi_i z^*_j$}
\begin{tabular}{|c|ccc|ccc|}
\hline
$\diagdown$ & $z_1$ & $z_2$ & $z_3$ & $z^*_1$ & $z^*_2$ & $z^*_3$\\
                    \hline
                    
 $\chi_1$   & $\frac{1}{2} z_2$&$\frac{1}{2} z_1$ & 0 &                                                                                              $-\frac{1}{2} z^*_2$ &$-\frac{1}{2} z^*_1$            & 0 \\ 
 $\chi_2$   &$\frac{1}{2} z_2$ &$-\frac{1}{2} z_1$ & 0 &$\frac{1}{2} z^*_2$&$-\frac{1}{2} z^*_1$&0\\
$\chi_3$   &$\frac{1}{2} z_1$ &$-\frac{1}{2} z_2$ & 0 &$-\frac{1}{2} z^*_1$&$\frac{1}{2} z^*_2$& 0  \\
&&&&&&\\
                     
 $\chi_4$   &$\frac{1}{2} z_3$ & 0 &$\frac{1}{2} z_1$ & $-\frac{1}{2} z^*_3$ & 0  & $-\frac{1}{2} z^*_1$  \\
 $\chi_5$   &$\frac{1}{2} z_3$ & 0 &$-\frac{1}{2} z_1$ & $\frac{1}{2} z^*_3$  & 0  & $-\frac{1}{2} z^*_1$  \\
 $\chi_6$   &0 &$\frac{1}{2} z_3$ &$\frac{1}{2} z_2$ & 0 & $-\frac{1}{2} z^*_3$  &  $-\frac{1}{2} z^*_2$   \\
 $\chi_7$   &0 &$\frac{1}{2} z_3$ &$-\frac{1}{2} z_2$ & 0 & $\frac{1}{2} z^*_3$  &$-\frac{1}{2} z^*_2$   \\
 $\chi_8$   &$\frac{1}{2\sqrt{3}} z_1$ &  $\frac{1}{2\sqrt{3}} z_2$  &$-\frac{1}{\sqrt{3}}z_1$  &$-\frac{1}{2\sqrt{3}} z^*_1$     & $-\frac{1}{2\sqrt{3}} z^*_2$  &$\frac{1}{\sqrt{3}} z^*_3$
 \label{chi-on-z}
 
\end{tabular}
\end{table*}

\begin{table*}[ht]
\caption{Gluon Multiplication Table}
\begin{tabular}{c| c c c c c c c c c}

$\diagdown$     &
 $\textit{R}\overline{R}$ & $\textit{R}\overline{G}$ & $\textit{R}\overline{B}$ &  $\textit{G}\overline{R}$ & $\textit{G}\overline{G}$ & $\textit{G}\overline{B}$ &  $\textit{B}\overline{R}$ & $\textit{B}\overline{G}$ & $\textit{B}\overline{B}$\\
 \hline
$\textit{R}\overline{R}$&$\textit{R}\overline{R}$&$\textit{R}\overline{G}$&$\textit{R}\overline{B}$& 0&0&0&0&0&0\\
$\textit{R}\overline{G}$&0&0&0&$\textit{R}\overline{R}$&$\textit{R}\overline{G}$&$\textit{R}\overline{B}$&0&0&0\\ 
$\textit{R}\overline{B}$&0&0&0&0&0&0&$\textit{R}\overline{R}$&$\textit{R}\overline{G}$&$\textit{R}\overline{B}$\\
\hline
$\textit{G}\overline{R}$&$\textit{G}\overline{R}$&$\textit{G}\overline{G}$&$\textit{G}\overline{B}$&0&0&0&0&0&0\\
$\textit{G}\overline{G}$&0&0&0&$\textit{G}\overline{R}$&$\textit{G}\overline{G}$&$\textit{G}\overline{B}$&0&0&0\\
$\textit{G}\overline{B}$&0&0&0&0&0&0&$\textit{G}\overline{R}$&$\textit{G}\overline{G}$&$\textit{G}\overline{B}$\\
$\textit{B}\overline{R}$&$\textit{B}\overline{R}$&$\textit{B}\overline{G}$&$\textit{B}\overline{B}$&0&0&0&0&0&0\\
$\textit{B}\overline{G}$&0&0&0&$\textit{B}\overline{R}$&$\textit{B}\overline{G}$&$\textit{B}\overline{B}$&0&0&0\\
$\textit{B}\overline{B}$&0&0&0&0&0&0&$\textit{B}\overline{R}$&$\textit{B}\overline{G}$&$\textit{B}\overline{B}$
\label{glu multip table}
\end{tabular}
\end{table*}

\begin{table*}[ht]
\caption{D(1,0) and D(0,1), Quark and Antiquark Triplets}
\begin{tabular}{c c c|c c c c c c}\\
~ & $f_{n_1n_2n_3}^{m_1m_2m_3}$ & $q,~\overline{q}$ &  $B$        & $I_3$             &  $Y$           &   $Q$            &  $S$  & $(4I^2+3Y^2)$\\
&&&&&&&\\
\hline\\
 ~      & $f_{100}^{000} $  &$u$     &   $\frac{1}{3}$  &  $+\frac{1}{2}$  & $+\frac{1}{3}$& $+\frac{2}{3}$ &  0  & 4/3\\
$[3]$ & $f_{010}^{000} $ & $d$     &   $\frac{1}{3}$  &  $-\frac{1}{2}$   & $+\frac{1}{3}$& $-\frac{1}{3}$  &  0 & 4/3 \\
 ~     &  $f_{001}^{000} $ & $s$     &  $\frac{1}{3}$  &  $ 0$                  & $-\frac{2}{3}$& $-\frac{1}{3}$   &  -1  & 4/3\\
 &&&&&&&\\
\hline\\
 ~&  $ f_{000}^{100} $ & $\overline{u}$ &  $-\frac{1}{3}$&  $-\frac{1}{2}$   & $-\frac{1}{3}$& $-\frac{2}{3}$ &  0 & 4/3 \\
$[\overline{3}]$ &  $f_{000}^{010} $ & $\overline{d}$ &$-\frac{1}{3}$  &  $+\frac{1}{2}$    & $-\frac{1}{3}$& $+\frac{1}{3}$  &  0  & 4/3\\
   ~ &  $ f_{000}^{001} $ & $\overline{s}$ &  $-\frac{1}{3}$&  $  0$            & $+\frac{2}{3}$ & $+\frac{1}{3}$  &  1 & 4/3 \\
\end{tabular}
%
\label{Table:Quark and Antiquark}
\end{table*}

\begin{table*}[ht]
\caption{D(1,1),  Pseudoscalar Meson Nonet}
\begin{tabular}{c c|c c c c c c}\\
 $f_{n_1n_2n_3}^{m_1m_2m_3}$   & Meson & $B$ & $I_3$ &  $Y$  &  $Q$ &   $S$   & $(4I^2+3Y^2)$ \\
\hline\\
 $f_{100}^{010}$ & $u\bar{d}$, $\pi^+$     &  0    &     1   &  0  &  1 &   0 & 4 \\
 $f_{010}^{100}$ & $d\bar{u}$, $\pi^-$      &  0    &    -1   &  0  & -1 &   0  & 4 \\

\hline\\
$f_{100}^{001}$ & $u\bar{s}, K^+$          &  0    & $\frac{1}{2}$   &  1  & 1 & 1   &4\\
$f_{010}^{001}$ & $d\bar{s}, K^0$           &  0    & $-\frac{1}{2}$ &  1  & 0  &  1   & 4\\
$f_{001}^{100}$ & $ s\bar{u}, K^-$           &  0    & $-\frac{1}{2}$ & -1  &-1  & -1   & 4\\
$f_{001}^{010}$ & $\bar{d} s, \bar{K}^0$  &  0    & $ \frac{1}{2}$ & -1  & 0  &   -1   & 4\\
\hline\\
$\frac{1}{\sqrt{2}}(f_{100}^{100}> - f_{010}^{010}>)$ &~ $\frac{1}{\sqrt{2}}(u\bar{u}- d\bar{d}),  ~~~~~~~~~ \pi^0$      &  0    &         0   &  0  &  0 &   0  & 0\\
$\frac{1}{\sqrt{6}}(f_{100}^{100}+f_{010}^{010}-2f_{001}^{001})$ &~ $\frac{1}{\sqrt{6}}  (u\bar{u}+ d\bar{d}-2s\bar{s})$,~ $\eta$       &  0    &  0             &  0  & 0  &   0  & 0 \\
$\frac{1}{\sqrt{3}}(f_{100}^{100}+f_{010}^{010}+f_{001}^{001})$ &~ $\frac{1}{\sqrt{3}}(u\bar{u}+ d\bar{d}+s\bar{s})$,~~ $\eta'$       &  0    &  0             &  0  & 0  &   0   & 0\\
\end{tabular}
\label{Table: Pseudoscalar Meson Nonet}
\end{table*}
\begin{table*}[ht]
\caption{D(3,0), Baryon Decuplet}

\begin{tabular}{c c|c c c c c c c}\\
 $f_{n_1n_2n_3}^{m_1m_2m_3}$ & $Baryon$&  $B$ & $I$ & $I_3$ &  $Y$ & $Q$ &  $S$ & $(4I^2+3Y^2)$\\
\hline\\
       $f_{120}^{000}$ & $udd, ~n$ ~          & 1  &  $\frac{1}{2}$ &$-\frac{1}{2}$& 1 & 0  &  0  & 4 \\
        $f_{210}^{000}$ & $uud,~ p$ ~          & 1  &  $\frac{1}{2}$ &$\frac{1}{2}$ & 1 & 1  &  0  & 4 \\
        $f_{021}^{000}$ & $dds, ~\Sigma^-$  &1   &1                      &-1                  & 0 &-1  &-1   & 4 \\
        $f_{201}^{000}$ & $uus, ~\Sigma^+$ &1   &1                      & 1                  & 0 & 1  &-1   & 4 \\
        $f_{111}^{000}$ & $uds, ~\Sigma^0$  &1   &1                      & 0                  & 0 & 0  &-1   & 4 \\
        $f_{012}^{000}$ & $dss, ~\Xi^-$         &1   &$\frac{1}{2}$    &-$\frac{1}{2}$ & -1& -1 & -2 & 4  \\
        $f_{102}^{000}$ & $uss, ~\Xi^0$         &1   &$\frac{1}{2}$   & $\frac{1}{2}$ & -1& 0  & -2 & 4  \\
\hline\\
       $f_{300}^{000}$ & $uuu, ~\Delta^{++}$ & 1 &$\frac{3}{2}$  & $\frac{3}{2}$ & 1 &   2 & 0  & 12\\
       $f_{030}^{000}$ & $ddd, ~\Delta^{-}$     & 1 &$\frac{3}{2}$  &-$\frac{3}{2}$& 1 & -1  & 0  & 12\\
       $f_{003}^{000}$ & $sss,  ~\Omega^-$     & 1  & 0                   & 0                   & -2& -1 & -3  & 12
\end{tabular}
%
\label{Table: Baryon Decuplet}
\end{table*}

\end{document}